\documentstyle[aps,preprint]{revtex}
\begin{document}
\draft
\preprint{IC/94/346; IMSc-94/44}
\title{Electromagnetic and Gravitational Scattering at Planckian
Energies}
\author{Saurya Das \footnote{E-Mail:~saurya@imsc.ernet.in}}
\address{The Institute of Mathematical Sciences, \\ CIT Campus,
Madras - 600 113,  India.}
\author{Parthasarathi Majumdar\footnote{E-Mail:~partha@ictp.trieste.it;
partha@imsc.ernet.in}}
\address{International Centre for Theoretical Physics, Trieste I-34100,
Italy\footnote{Permanent address (after December 2, 1994): The Institute of
Mathematical Sciences, Madras 600 113, India.}}
\maketitle
\noindent
\begin{abstract}
The scattering of pointlike particles at very large center of mass
energies and fixed low momentum transfers, occurring due to both their
electromagnetic and gravitational interactions is re-examined in the
particular case when one of the particles carries magnetic charge.
At Planckian center-of-mass energies, when
gravitational dominance is normally expected, the presence of magnetic
charge is
shown to produce dramatic modifications to the scattering cross section
as well as to the holomorphic structure of the scattering amplitude.

 \end{abstract}

\narrowtext
\section{INTRODUCTION}
\label{sec:intro}

In perturbative quantum field theory , the entire
information about interactions is customarily relegated to the perturbing
Hamiltonian, with the exactly integrable part corresponding
to free propagation of quanta (forward scattering).  In situations where
a well-defined perturbative domain is not available, such a decomposition
of the original Hamiltonian into  `free' and `interacting' parts is
clearly not meaningful. It is more desirable, therefore, to formulate the
theory in
such a way that the part that is exactly tractable contains non-trivial
information about the interaction, even though this may be
semi-classical. In some cases, it turns out that there are kinematical
regimes where in
fact the semi-classical approximation is exact, permitting calculation
of scattering amplitudes without further approximations. We shall focus
on two such cases in the sequel. The first deals with an electromagnetic
system consisting of a point charge and a Dirac monopole, both of very
small mass. The second is basically a generalization of the first, in
which gravitational interactions between these particles is also taken
into account.

It is well known that a local field theory of electromagnetism incorporating
{\it both} electric and magnetic charges is not as easy to formulate as one
with electric charges alone. Furthermore, if we assume that the electric
charge is small, given essentially in terms of the fine structure
constant, then the magnetic charge, by virtue of Dirac quantization will
certainly not be small. Thus, the sector of the theory with magnetic charge is
not amenable to a perturbative treatment. However, there exists a
kinematical region in which exact computation of the scattering amplitude
of these particles is possible. The way this comes about
is the following: if we imagine a situation in which the center of mass
(c.m.) energy of the system is very high, while the momentum transfer
between the scattering constituents is fixed at a relatively low
value, then many of the degrees of freedom of the system
decouple. The remaining degrees of freedom
become strongly coupled and turn out to be accessible to exact
analyses without further approximations. In the case of pure electric
charge-charge scattering, the amplitude corresponds exactly to the one
calculated in the so-called Eikonal approximation of quantum
electrodynamics. In this case, of course, radiative corrections can be
calculated perturbatively, unlike in the charge-monopole case.

When the c.m. energies approach Planckian values, quantum effects of
general relativity can no longer be ignored. But, as of now, there is no
fully satisfactory
quantized theory of gravity. When one tries to quantize gravity
from a local field theoretic viewpoint, one immediately runs into
uncontrollable ultraviolet divergences. The string theory approach,
though excellent from the standpoint of perturbation theory, is yet
to be completely understood on a non-perturbative basis. Other
approaches like the Ashtekar formalism are not developed
well enough for analyzing physical processes involving exchange
of gravitational quanta. However, as we shall see, in the kinematical
regime under consideration, amplitudes of several
processes involving gravitational interactions
become exactly calculable, despite the lack of a full quantum gravity
theory. Furthermore, the interplay between gravitational and
electromagnetic interactions become especially interesting in this
kinematical regime when one of the particles is magnetically charged.
In this case the fine structure constant
of electromagnetism $\alpha$
does not evolve with the c.m. squared energy $s$, but increases with
increasing the squared momentum transfer
$t$.   So if $t$ is held fixed then $\alpha$ does not run at all. Thus,
in the kinematical region of interest, one expects gravitational
interactions to dominate over electromagnetism. With monopole-charge
scattering, though, this is not the case as we show below.

We shall see that longitudinal and
transverse degrees of freedom behave quite differently in the
above
situation, with the latter essentially
dropping out of the problem!
This will lead to the truncation of the full action of the theory
under consideration (both for general relativity and quantum
electrodynamics) to a two dimensional action defined on the
boundary of space time. In this sense, the theory has a distinctly
topological nature and yet non-trivial dynamical results follow from it.

The paper is organized as follows. In the second section we review
earlier literature on pure electric charge-charge scattering within the
``shock-wave'' picture. Scaling arguments leading to a truncation of the
Maxwell action and eventually to the shock wave picture will be
summarized for completeness. Then we introduce magnetic monopoles in the
theory,
and proceed to generalize the foregoing formalism to calculate the
scattering amplitude. Particular attention will be paid to subtleties
arising from problems like the Dirac string singularity. In the third
section gravity will be introduced and the
interactions involving both electromagnetism and gravity will be
studied. Once again we will motivate the discussion by considering the
full
Einstein action and how it gets simplified, in the absence of
electromagnetism \cite{ver2}.  Next we discuss charge-charge and
charge-monopole scattering at
Planckian energies. The relative contributions of electromagnetic and
gravitational scattering in the two cases will be contrasted in detail. We
will also comment on
the behavior of singularities, namely the poles in the scattering
amplitude and how they differ from one process to another. We conclude
with a number of observations on our results and future outlook.

\section{Electromagnetic scattering at high energies}
\label{sec:em}

At sub-Planckian c.m. energies that are still large compared to the rest
masses of the particles, the dominant physical processes originate from a
truncated version of the original Maxwell action. The derivation of this
truncation is first briefly sketched, and the resulting shock wave fields
calculated in a frame where one of the particles moves almost luminally.
The other scattering particle, assumed to be relatively slow, scatters
off these fields with an exactly computable amplitude. The review of this
material follows the treatment of Verlinde and Verlinde \cite{ver} and of
Jackiw et. al. \cite{jac}, and is followed by generalization to the case
of monopole-charge scattering.

\subsection{Effective Theory at High Energies}
\label{subsection:scale}

Suppose there are two spinless
charged particles moving at very high velocities, such that the
center of mass energy $\sqrt s$ is very high.  The action
for the electromagnetic field is given by
\begin{equation}
S~=~-{1 \over 4}\int d^4x \left ( F_{\mu \nu}F^{\mu \nu} \right )
\label{action}
\end{equation}
where
$F_{\mu \nu}~=~{\partial}_{\mu}A_{\nu} -
{\partial}_{\nu}A_{\mu},$ is the second rank electromagnetic
field strength tensor and $A^{\mu}=\left(
A^{0}, \vec A \right)$ is the electromagnetic four potential.
At high center of mass energies and very low momentum transfer
$\sqrt t$, the scattering is almost exclusively in the forward
direction. Without loss of generality, if we assume the particles to move
initially in
the $z$ direction,
with $4$-momentum $p^{\mu}=(p^{0}, \vec p)$, then we have
for lightlike particles, the energy, $E \approx p_z \sim \sqrt s$
and $p_x=p_y \sim 0$. The square roots of $s$ and $t$ thus measure the
typical momenta
associated with the longitudinal and the transverse directions.
Now, if we associate two length scales with the longitudinal and
transverse directions, then
the characteristic transverse length scale is much bigger than the
longitudinal length
scale. Thus, we scale the null coordinates $x^{\pm}$
such that $x^\alpha \rightarrow \lambda
x^\alpha$ and $x^{i} \rightarrow x^{i}$, where
$\alpha$ runs over the
light cone indices $+,-$, while $i$ signifies the transverse
coordinates $x,y$. Under this scaling the $A_{\mu}~s$ transform as
$A_\alpha \rightarrow
{\lambda}^{-1}A_{\alpha}$. The transverse $A_i~s$ remain unchanged.
The transformed action now has the form
\begin{equation}
S~=~-{1 \over 4}\int d^4x \left( {\lambda}^{-2}F_{\alpha \beta}F^{\alpha
\beta}~+~2F_{\alpha i}F^{\alpha i}~+~
{\lambda}^2 F_{ij}F^{ij} \right) ~.
\end{equation}

The parameter $\lambda$ may now be chosen to depend on $s$ :
\begin{equation}
\lambda = {k \over {\sqrt s}}\rightarrow 0~~,
\end{equation}
where $k$ is a finite constant having dimensions of energy.
Then the limit $s \rightarrow \infty$ becomes equivalent to the
limit $\lambda \rightarrow 0$. Thus in this kinematical regime,
the transverse part of the action with $F_{ij}$
can be ignored and what we have
left is an effective action of the form
\begin{equation}
S~=~-{1 \over 4}\int d^4x \left( {\lambda}^{-2}F_{\alpha \beta}F^{\alpha
\beta}~+~2F_{\alpha i}F^{\alpha i}\right)
\end{equation}
Notice that in the partition function the fluctuations of the
term $F_{\alpha \beta}F^{\alpha \beta}$ are suppressed
in the imaginary exponent (due to
the smallness of $\lambda$) and the configuration with the dominant
contribution is
$F_{\alpha \beta}=0$ i.e. $F^{+-}=E_z=0$ \cite{li}.
This shows that the electric field is localized in the transverse
plane. Similarly, if we write the original action in the dual
formalism, with the $F_{\mu \nu} \rightarrow {\tilde F}_{\mu \nu}$, then
${\tilde F}^{+-}=B_z=0$. This brings us to the {\it shock wave picture}:
fields due to processes characterized by longitudinal momenta that are
overwhelmingly larger than transverse momenta are essentially
confined to the plane (called the `shock front')
perpendicular to the direction of motion of the source particles.

{}From the field theory standpoint, a charged scalar field theory coupled
to electromagnetism also undergoes simplification in this kinematical
regime : the action under the same scale transformation, becomes
\begin{equation}
S~=~\int d^4x \left( D_{\alpha}{\phi}D^{\alpha}{\phi}^{*} +
{\lambda}^2 D_{i}{\phi}D^{i}{\phi}^{*} \right)
\end{equation}
Once again on neglecting terms of order ${\lambda}^2$,
we see that only the longitudinal components of the gauge
fields remain coupled. Thus, if we were to describe the gauge field
interaction in terms of currents $j^{\mu}$, then only the light cone
components $j^{\pm}$ would be physically relevant. Furthermore, if these
currents were to be associated with charges moving almost
luminally, then
 \begin{equation}
j_{\pm}~=~j_{\pm} \left( x^{\pm}, {\vec r}_{\perp} \right) ~~~~
j^{i}(x)~=~0
\end{equation}
This allows us to define two functions $k^{+}$ and $k^{-}$, where
\begin{eqnarray}
\label{eq:source}
j_{+}~=~{\partial}_{-} k^{-} \left( x^{-}, {\vec r}_{\perp} \right) \nonumber
\\
j_{-}~=~{\partial}_{+} k^{+} \left( x^{+}, {\vec r}_{\perp} \right).
\end{eqnarray}
In short, if we define a vector $k$ such that
\begin{equation}
k\left(x\right)~=~k^{+}\left(x^{+}, {\vec r}_{\perp} \right) -
k^{-}\left(x^{-}, {\vec r}_{\perp} \right)
\end{equation}
then
\begin{equation}
j^{\alpha}~=~{\epsilon}^{\alpha \beta}{\partial}_{\beta}k,
\end{equation}
where ${\epsilon}^{\alpha \beta}$ is antisymmetric and
${\epsilon}^{01}=1$. The above form of $j^{\alpha}$
automatically ensures
the current conservation ${\partial}_{\alpha}j^{\alpha}=0$.

The flatness condition $F^{+-}=0$ above admits a solution in terms of the
light cone components of the gauge
potential $A_{\pm}={\partial}_{\pm}{\Omega}$. If, further, we impose
the Landau gauge ${\partial}_{\mu}A^{\mu}=0$, $\Omega$ obeys d`Alembert's
equation
 \begin{equation}
{\partial}_{+}{\partial}_{-}{\Omega}~=~0
\end{equation}
which implies
\begin{equation}
\Omega~=~{\Omega}^{+}\left(x^{+}, {\vec r}_{\perp} \right) +
{\Omega}^{-}\left(x^{-}, {\vec r}_{\perp} \right).
\end{equation}
It is then easy to show that the electromagnetic
Lagrange density $${\cal L}~=~ -{1 \over 4}F_{\mu \nu}F^{\mu
\nu} - j^{\mu}A_{\mu}$$ can be written as
\begin{equation}
{\cal L}~=~-{1 \over 2}{\partial}_{-}{\Omega}^{-}{\vec
\nabla}^2{\partial}_{+}{\Omega}^{+}~-~
{1 \over 2}{\partial}_{+}{\Omega}^{+}{\vec
\nabla}^2{\partial}_{-}{\Omega}^{-}
- \partial_+k^+ \partial_-\Omega^-
- \partial_-k^- \partial_+\Omega^+
\end{equation}
which reduces to a total derivative  in the light cone coordinates
\begin{equation}
{\cal L}~=~-{\partial}_{-} \left( {1 \over 2}{\Omega}^{-}~{\vec
\nabla}^2 {\partial}_{+}{\Omega}^{+} +
{\partial}_{+}k^{+}~{\Omega}^{-} \right ) - {\partial}_{+} \left( {1
\over 2} {\Omega}^{+}~{\vec \nabla}^2 {\partial}_{-} {\Omega}^{-}
+ {\partial}_{-}k^{-} {\Omega}^{+} \right).
\end{equation}
This shows that the action $S=\int d^4 x {\cal L}$ is a surface term
defined on the boundary of null plane:
\begin{equation}
S~=~\oint~d\tau \int~d^2r_{\perp}~ \left( {1 \over 2}
{\Omega}^{-}~{\vec \nabla}^2 {\dot {\Omega}}^{+} - {1 \over
2}{\Omega}^{+}~{\vec \nabla}^2{\dot {\Omega}}^{-} + {\dot k}^{+}
{\Omega}^{-} - {\dot k}^{-} {\Omega}^{+} \right) .
\end{equation}
Here all the quantities are evaluated on the contour
parametrized by the affine parameter $\tau$. An overdot
denotes ${\partial}/{\partial}\tau$.
This shows that although the Lagrange density was reduced
to a total derivative, the values of the gauge parameter at the
boundary plays a significant role. In fact, they are the only
dynamical degrees of freedom in the problem. This simplification
of the action has its origin in the kinematics of the situation.
On extremising this action, the equations of motion obtained are
\begin{eqnarray}
{\nabla}^2{\dot \Omega}^{+}~&=&~-{\dot k}^{+}  \nonumber \\
{\nabla}^2{\dot \Omega}^{-}~&=&~-{\dot k}^{-},  \\
\end{eqnarray}
which yields on integration
\begin{eqnarray}
{\Omega}^{+} \left(x^+,{\vec r}_{\bot} \right)~&=&~-{1 \over
{\nabla}^2}~k^+ \left(x^+, {\vec r}_{\bot} \right)    \nonumber  \\
{\Omega}^{-} \left(x^-,{\vec r}_{\bot} \right)~&=&~-{1 \over
{\nabla}^2}~k^- \left(x^-, {\vec r}_{\bot} \right) .
\end{eqnarray}
It can be verified that these solutions are identical to those obained
by solving the full set of Maxwell's equations with (\ref{eq:source}) as
the source current. In other words, once again we arrive at the shock
wave description of highly energetic charged particles.
It can be shown \cite{jac} that exact scattering amplitude
for charge-charge scattering, to be computed below semiclassically,  can
also be obtained  from the above reduced action.

\subsection{Charge-charge scattering}
\label{subsection:charge}
The foregoing analysis allows us to compute {\it exactly} the $S$-matrix
for the scattering of two highly energetic particles assumed to  carry
electric charge.  Making use of Lorentz covariance of the theory, we
will do the calculations in a special inertial frame in which one
of the charges moves with velocity close to luminal, while the other is
moving relatively slowly. The shock wave front due to the former extends
over the entire transverse plane. Thus, the target particle, assumed to be
moving in a direction opposite to that of the source, encounters this
shock wave and its wave function acquires an Aharanov-Bohm type phase
factor. The overlap between the wave functions of the target particle
before and after its encounter with the shock front leads to the
scattering amplitude.

The potential of the lightlike particle can be found in
various ways. First, we approach it from a very well known physical
situation, namely that of Cerenkov radiation. If a
particle carrying an electric charge $e'$
moves in the positive $z$ direction in a
dielectric medium with a dielectric constant
$\epsilon$, at a speed $\beta$ greater
than the speed of light in that
medium, then it emits electromagnetic radiation. The charge
carries with it a shock wave, in front of which all
potentials and fields vanish. The vector potential due to this
charge behind the shock wave is given by the formula (e.g.
\cite{jackson})
\begin{equation}
A_z(x,y,z)~=~ {{\beta}e' \over {\sqrt {{\left( z- \beta t \right)}^{2}
 + {\left( 1 - {\beta}^{2}\epsilon \right)}~{r_{\perp}}^2}}},
\end{equation}
$A_x$ and $A_y$ being zero. $r_{\perp}$ is the transverse
distance from the charge given by
${r_{\perp}}^{2}=x^{2}+y^{2}$.
Thus $\vec A$ suffers a discontinuity across the shock front
giving rise to singular fields. Now if we put
$\epsilon=1$,
which means that the motion is in vacuum, and take the limit
$\beta \rightarrow 1$, then
expression for $\vec A$ will be the quantity of our interest.
Of course, now the charge will move exactly at the speed of light
in vacuum.

The same result can be derived somewhat more formally following \cite{jac}.
We consider the electromagnetic
potential $A_\mu$ of a static charged particle
with an electric charge $e'$. Then we give it a Lorentz boost
$\beta $ along the positive $z$ axis. The gauge potential
transforms
accordingly following the laws of special relativity. On taking
the limit $\beta \rightarrow 1$, the potential of the lightlike
particle is found to be a pure gauge almost everywhere except on
the shock plane where it has a discontinuity,
\begin{eqnarray}
\tilde A^{0}&=&\tilde A^{z}=-2e'~ \ln (\mu r_{\perp})~ \delta(x^{-})
, \nonumber \\
{\tilde A}_{\perp}^{i}&=&0,~~~~~~~~ i=1,2.
\label{eq:pot}
\end{eqnarray}
Here, $\mu$ is a dimensional parameter inserted to make the logarithm in
\ref{eq:pot} dimensionless. The potential $\tilde A^{\mu}$ is singular on
the shock
plane $(x^{-}=0)$ as was seen from the Cerenkov formula. Now this
potential is gauge equivalent to the potential
$A^{'}{}^{\mu}$ where $A^{'}{}^{\mu}=\tilde A^{\mu}+\partial ^{\mu}\Lambda,
{}~\Lambda $ being a Lorentz scalar. Choosing $\Lambda$ to be
$-2e' \theta(x^-) \ln \mu r_{\perp}~ $, we get
\begin{equation}
A^{'}{}^{0}=A^{'}{}^{3}=0,~~~~~
\vec{A}^{'}{}_{\perp}=-2e^{'}\theta (x^{-})
\vec \nabla \ln \mu r_{\perp}
\label{eq:atotal}
\end{equation}
We see that the gauged transformed vector potential is a pure
gauge everywhere except on the hyperplane $x^{-}=0$ which is also
the shock plane. Thus as one expects, the fields are
non-vanishing only on this plane and are given by
\begin{eqnarray}
E^i &=&  {2e^{'} r_{\perp}^i \over {r_{\perp}^2}}{\delta
(x^{-})},~~~~~~~~~~ E^z=0  \nonumber \\
B^i &=& -{2e^{'} \epsilon_{ij} r_{\perp}^j \over
r_{\perp}^2}{\delta
 (x^{-})} ,~~~~ B^z=0~.
\label{eq:efield}
\end{eqnarray}
These singular field configurations cause an instantaneous interaction
with the (slower) target particle.  First, consider the classical
motion of the slow particle. This reduces to solving the Lorentz
force equation for the test charge $e$ of mass $m$  with given boundary
conditions.  Since it has negligible velocity, we use the non-relativistic
form of the equation and also neglect the $\vec B$ dependent
piece. Thus we have
\begin{equation}
m{d^2 \vec r \over dt^2}~=~{2ee^{'} \over r^2_\perp} \delta(t-z)
{}~{\vec r}_{\perp}
\label{eq:lorentz}
\end{equation}
The solution to the above equation can be easily guessed. Without
loss of generality, let us assume that at the initial time
$t=0$, ~$e$ is almost stationary on the $x$ axis at a distance $b$
from the origin.
As the electric fields are all directed radially, the
impulse imparted to $e$ should be along the positive $x$ axis
after which it starts moving in that direction with a uniform
velocity. The delta function shows that the shock wave arrives from the left
and hits it at $t=0$. Being non zero only at that instant, it also allows us
to replace $x$ and $y$ by $b$ and $0$ respectively on the right of the
equation. Inserting the constants correctly we have the solution
\begin{equation}
y(t)~=~z(t)~=~0,~~~~~~~x(t)~=~{2ee^{'} \over bm} t~\theta (t) + b
\end{equation}
which clearly satisfies equation (\ref{eq:lorentz}).
This is the classical trajectory of the charge $e$. The total momentum
transfer (or the impulse) is
just $2ee^{'} / b$ which is finite although the fields are
singular on the shock plane.

Having solved the classical part,
we now consider the quantum problem for the charge $e$.  As stated
earlier, we look at how the wavefunction changes under
the influence of the other charge which effectively provides just
a classical background field.
For early times $t<z$ the particle is free and its
wavefunction is just a plane wave given by,
\begin{equation}
\psi _<~(x^{\pm}, {\vec r}_{\perp}) =~\psi _0~=~\exp[ipx]~~~~for~ x^-<0~.
\end{equation}
with momentum eigenvalue $p^{\mu}$.
Immediately after the shock front passes by, its interaction with
the gauge potential enters via the `minimal coupling
prescription' by which we replace all the $\partial_{\mu}$'s by
$\partial_{\mu}-ieA_{\mu}$. The corresponding wavefunction
acquires a multiplicative phase factor $exp\left(ie\int dx^{\mu}A_{\mu}
\right)$.
Thus from equation (\ref{eq:atotal}), for $x^{-}>0$,
the modified wavefunction is
\begin{equation}
\psi _>(x^{\pm}, {\vec r}_{\perp})~ =~ \exp~[-iee^{'}
\ln(\mu^{2}r^{2}_{\perp})]~\psi '_0
{}~~~~for~ x^->0
\label{eq:psi}
\end{equation}
where $\psi_0$ and $\psi'_0$ are related through the continuity
requirement
\begin{equation}
\psi _< ~=~ \psi _> ~~~~ at~ x^- = 0 .
\end{equation}
Here it may be noted that the additional phase factor due to
electromagnetic interaction is a function of $r_\perp$ only,
which is the length of the radius vector from the particle on the
shock plane. It does not depend on the angular variable. This is
due to the fact that the electric field of an
electrically charged particle is central in nature.
The wavefunction $\psi_>$ can now be expanded in terms of
the complete set of
momentum eigenfunctions (plane waves) with suitable coefficients
in the following form \cite{thf}
\begin{equation}
\psi _> = \int dk_+d^2k_{\bot}~A(k_+ , {\vec k}_{\bot}
)\exp[i{\vec k}_{\bot} \cdot {\vec r}_\bot -ik_+x^- -
ik_-x^+ ] \label{eq:expn}
\end{equation}
with the on shell condition $k_+ = {(k^2_{\perp} +
m^2)/k_-}~.$ Obviously the coefficients $A(k_+,k_{\perp})$ are the
probability amplitudes for finding the particle with momentum
$k^{\mu}$ when an experiment is performed on it after it has
undergone the shock wave interaction. So we proceed to calculate
them by
multiplying both sides of equation (\ref{eq:expn}) by a
plane wave and integrating
over $x^-$. Using the orthonormality of the eigenfunctions,
we get
\begin{equation}
A(k_+,k_\bot) = {\delta (k_+ - p_+)
\over (2 \pi)^2}{\int
d^2r_{\bot} \exp i \left(-2ee^{'}~\ln(\mu r_\perp) +
{\vec q} \cdot {\vec r}_\bot \right)}~,
\end{equation}
where ${\vec q} \equiv
{\vec p}_{\bot} - {\vec
 k}_{\bot}$ is the
transverse momentum
transfer, $k$ and $p$ being the final and initial momenta
respectively.  The integration over the transverse $x-y$ plane can be
performed exactly \cite{jac} yielding the amplitude
\begin{equation}
f(s,t)= {k_+ \over 4\pi k_0}~{\delta(k_+ - p_+)}
{\Gamma
(1-iee') \over \Gamma (
iee')}\left (4 \over
-t \right
 )^{1-iee'}~~.
 \label{eq:gram}
 \end{equation}
where we have put in the canonical kinematical factors. $t \equiv
-q^2$ is the transverse momentum transfer. With this
amplitude, one can easily show that the scattering cross section is
\begin{equation}
{d^2\sigma \over d\vec k_{\perp}^2}~ \sim~ {(ee^{'})^2 \over t^2 }~~,
\end{equation}
where we have used a property of the gamma function, namely
${|\Gamma (a+ib)|}={|\Gamma (a-ib)|}$, $a$ and $b$ being real.

It has been  shown in \cite{jac} that this scattering amplitude is
identical to the amplitude obtained in the Eikonal approximation where
virtual momenta of exchanged quanta are ignored in comparison to external
momenta, leading to a resummation of a class of Feynman graphs
\cite{arb}. Since, generically $ee'={\cal O}({1 \over 137})$, this
approximation will receive usual perturbative radiative corrections.
The second order pole singularity in
the cross section as $t \rightarrow 0$ is, of course, typical of processes
where massless quanta are exchanged.

\subsection{Charge - monopole scattering}

Now that we have calculated the amplitude of the
scattering of two charges, one can inquire as to  what
changes, if any, will take place if we replace one of the charges
by a Dirac magnetic monopole. This question is worth pursuing for various
reasons. First of all, the (albeit imagined) existence of monopoles will
imply that the Maxwell equations assume a more symmetric form, due to the
property of duality of field strengths and electric and magnetic charges.
Within quantum mechanics, as Dirac has shown, monopoles offer a
unique explanation of the quantized nature of electric
charge. But as is well-known, introduction of monopoles in the
theory brings in other problems such as singularities in the
vector potential. It will be interesting to see how one can deal
with them in the present formalism and investigate the range of validity
of the shock wave picture in this context. One should also keep in mind
the fact that a satisfactory local quantum field theory for monopoles is
still lacking. Further, given Dirac's quantization condition, monopole
elelctrodynamics cannot be understood in perturbative terms around some
non-interacting situation. Thus, as advertized earlier, the shock wave
picture may be one of the few important probes available for such processes.

Recall, however, following \cite{wu} that it is not
possible to choose a single
non-singular potential to describe the field of the monopole everywhere.
We need at least two such
potentials, each being well behaved in some region and being related by a
local gauge transformation in the overlapping region.  In spherical polar
coordinates, these potentials can be chosen as \cite{wu},
\begin{eqnarray}
{\vec {A}}~^I ~& = &~ {g \over {r \sin \theta} } (1 - \cos
\theta) {\hat \phi}~,~~~~~  0 \le \theta < \pi
\nonumber  \\
{\vec{A}}~^{II}~& = &~{-~g \over { r \sin \theta}}
(1+\cos\theta){\hat \phi}~.~~~~~  0 < \theta \le
\pi~.\label{eq:vec2}
\end{eqnarray}
The Dirac strings associated with the two potentials are along
the semi infinite lines $\theta= \pi$ and $0$ respectively,
i.e. along the negative and positive halves of the $z$
axis. ${\vec A}^{I}$ and ${\vec A}^{II}$ become singular along these two
lines respectively.
It may be noted that here we have made the gauge choice
$A^0=0$, and have chosen an orientation of our coordinates such
that only the $x$ and $y$
components survive. In the region $-\pi < \phi < \pi$, where either of
$\vec {A}~^{I}$ or $\vec {A}~^{II}$ may be used, they are related
by a gauge transformation with the gauge parameter $2g\phi$. It
can be readily verified that
\begin{equation}
\vec \nabla  \times \vec{A}~^{I} ~=~ \vec \nabla \times
\vec{A}~^{II} ~=~{g \over r^2}\hat{r}.
\end{equation}
Here the curls are taken in the respective regions of
validity of the potentials. In the following calculations for
convenience we shall work with ${\vec A}~^{I}$ only,
but all subsequent results will be independent of this particular choice.

As in the last section, we give the monopole a Lorentz
boost of magnitude $\beta$ along the positive z axis.
It can be shown that
if equations (\ref{eq:vec2}) are rewritten in cartesian coordinates,
then
$\vec{A}~^{I}$ transforms to \cite{das}
\begin{equation}
{}^{\beta}{A}^I_i={-g \epsilon_{ij} r^j_{\perp} \over
r^2_{\perp}
 }\left [1 - {{z
 - \beta t} \over R_{\beta}} \right ] ~, ~
\label{eq:veb}
\end{equation}
Before proceeding further, let us examine the behavior of the Dirac
strings under Lorentz boosts. For this purpose it is convenient to write
equation (\ref{eq:veb}) in the following form.
\begin{equation}
^{\beta}{\vec A}^{I}~=~{g \over r_{\bot}} \left[ 1 - {{z - \beta t}
\over R_{\beta}} \right] \hat \phi
\label{eq:str}
\end{equation}
On the $z$- axis $\left( \theta=0~~ or~~ \pi \right)$, we have $ r_{\bot}
\rightarrow 0$ implying that $R_{\beta} \rightarrow |z - \beta t|$. Thus
the above equation reduces to
\begin{equation}
^{\beta}{\vec A}^{I}~=~{g \over r_{\bot}} \left[ 1 - sgn (z - \beta t)
\right] .
\end{equation}
Thus for $z>\beta t$, i.e. in front of the boosted monopole the vector
potential vanishes, while it becomes singular behind it $\left( z <
\beta t \right)$. It is as if the monopole drags the Dirac string along
with it and as in the static case, the semi - infinite line of
singularity originates from it. Similarly, by looking at the boosted
potential $\vec A^{II}$, it can be easily verified that for this, the
string is always in front of the monopole and `pushed' by it as it moves.
These results also hold in the limit $\beta \rightarrow 1$, i.e. for the
potential,
\begin{equation}
\vec{\tilde A}~^I_0~\equiv \lim_{\beta \rightarrow 1}~
^{\beta}{\vec A}_i~^I = {2g \over r_{\perp}}~\theta(x^-) {\hat \phi}.
\label{eq:boos} \end{equation}
The corresponding electromagnetic fields are
\begin{eqnarray}
B^i ~&=&~  {2g r_{\perp}^i \over {r_{\perp}^2}}{\delta
(x^{-})},~~B^z=0  \nonumber \\
E^i ~&=&~ {2g \epsilon_{ij} r_{\perp}^j \over
r_{\perp}^2}{\delta
(x^{-})} ,~~ E^z=0~. \label{eq:fld}
\end{eqnarray}
Unlike the fields of a charge in motion, here the magnetic field
is radial, whereas the electric field is circular on the shock
plane. Here also $\vec A^{I}_{0}$ is a pure gauge everywhere except on
the null plane $x^{-}=0$. It may be noted that the above $\vec E$
and $\vec B$ fields can be obtained by making the following
transformations in (\ref{eq:efield}):
$e^{'} \rightarrow g$, $\vec E
\rightarrow \vec B$ and $\vec B \rightarrow -\vec E$. This is a
consequence of the duality symmetry in Maxwell's equations
incorporating monopoles.

As before let us now calculate the classical trajectory of the charge
under the influence of the monopole shock wave.
Here the non-relativistic Lorentz force equation for $e$ becomes
\begin{equation}
m{d^2 \vec r \over dt^2}~=~{2eg \over r^2_\perp} \delta(t-z)
{}~[\hat x y - \hat y x]
\end{equation}
where we have ignored the velocity of the slow test charge.
Imposing identical boundary conditions for the charged particle $e$ as
before and taking into account the fact that in this case the momentum
transfer will be along the $y$ axis (now that the $\vec E$ field lines
are circles on the shock plane) the solution $\vec r(t)$ is
\begin{equation}
x(t)~=~b~,~~~~z(t)~=~0~,~~~~y(t)={2eg \over bm}t \theta(t).
\end{equation}
In this case the impulse is $2eg/bm$.

In the quantum case, the charge $e$ interacts instantaneously with the
monopole shock wave, the net effect being a
gauge rotation in the wavefunction of the former.
To compute this explicitly, we proceed as follows \cite{das}. We first
rewrite
${\vec {\tilde A}}~^I_0$ in (\ref{eq:boos}) as a total derivative in the
following form
\begin{equation}
\vec {\tilde A}~^I_0~ =~ 2g \theta (x^{-}) \vec {\nabla} \phi ~.
\label{eq:dif} \end{equation}
We note in passing that the gauge potentials for a luminally boosted
electric charge (\ref{eq:atotal}) and monopole (\ref{eq:dif}), both given
as total derivatives on the transverse plane, form the real and imaginary
parts respectively of the gradient of the holomorphic function $lnz$
where $z \equiv r_{\perp} e^{i \phi}$, where $\phi$ is now the azimuthal
angle on the transverse plane.

For $t<z$, i.e. before the arrival  of the monopole with its shock front, the
wave function of the charge $e$ is once again the plane wave
\begin{equation}
\psi _<(x^{\pm},{\vec r}_{\perp})~ =~\psi _0 ~~~~for~ x^-<0~. \label{bef}
\end{equation}
After encountering the shock wave, it is modified by the
gauge potential dependent phase factor. The final form of the
wave function is
\begin{equation}
\psi _>(x^{\pm},{\vec r}_{\perp})~ =~ exp[i2eg \phi]~\psi '_0  ~~~~for~ x^->0
\end{equation}
by virtue of the potential (\ref{eq:dif}) with the usual
requirement of continuity. At this point we make the additional
assumption of Dirac
quantization namely, for an interacting monopole-charge
system, the magnitudes of their electric and magnetic charge must be
constrained by the relation
\begin{equation}
e~g~=~{n \over 2},~~~~~n~=~0,\pm1,\pm2,....
\end{equation}
Thus we get
\begin{equation}
\psi_>~=~e^{in\phi}{\psi}^{'}_{0}~.
\end{equation}
This sort of phase factor in the small angle scattering of a
monopole and a charge was first found by Goldhaber \cite{gold}.
It depends on the angular variable ${\phi}$ only.  This may be a
reflection of the non-central nature of the classical charge-monopole
interaction.

Expanding $\psi_>$ in plane waves as before we get an integral
expression for the scattering amplitude as follows
\begin{equation}
A(k_+,k_\bot) = {\delta (k_+ - p_+) \over (2 \pi)^2}{\int
d^2r_{\bot} \exp i\left(n\phi + {\vec q} \cdot {\vec r}_\bot
\right)}~.
\label{amp}
\end{equation}
Once again $\vec q \equiv p_\perp - k_\perp$ is the momentum
transfer and as before we have the dispersion relation $k_+ =
{(k^2_{\bot} + m^2)/k_-}~.$
By conveniently choosing the orientation of the transverse
axes as in the previous section, the angular integration
gives
$(1 / q^2) \int_{0}^{\infty} d\rho ~ \rho J_n(\rho)~,$
where $J_n(\rho)$ is the Bessel function of order $n$. This
integral is also standard \cite{luke} and the result is
\begin{equation}
\left ({1\over -t}\right )  {2\Gamma (1+ {n \over 2}) \over
\Gamma ({n
\over 2})} ~,
\label{rad}
\end{equation}
Here we note an important difference with the previously
calculated charge-charge amplitude. There the arguments of the
gamma functions were complex, whereas in this case they are
real. In fact, the amplitude in this case is simply
\begin{equation}
f(s,t)~=~{k_+ \over 2\pi k_0}~{\delta (k_+ - p_+) \left ({n \over
-t}\right)}~, \label{eq:scam}
\end{equation}
where we have incorporated the canonical kinematical factors.
Such factorization makes the expression for the amplitude simple. We
observe that it is proportional to the
monopole strength $n$. It follows that the scattering cross
section becomes
\begin{equation}
{d^2\sigma \over d\vec k_{\perp}^2}~ \sim~ {n^2 \over t^2}
\end{equation}

It may be mentioned that we would have obtained the same result if we
had used the second of the gauge potentials in
(\ref{eq:vec2}) and performed the Lorentz boost etc.  One way to see this
is by noting that the potentials, boosted to $\beta \approx 1$, are both
gauge
equivalent to a gauge potential $A'_{\mu}$ given by
\begin{equation}
{\vec A}'_{\perp}~=~0~=~A'_+~;~~~A'_-~=~2g \phi \delta(x^-)~~everywhere ~.
\end{equation}
The apparent disappearance of the Dirac string singularity in this
gauge is a red herring;  the gauge transformation has flipped the
Dirac string onto the shock plane, thus preventing it from being
manifest. More importantly, the gauge potential, though globally defined
functionally, is not single-valued, being a monotonic function of a
periodic angular variable. Thus, the singularity has been traded in for
non-single-valuedness. Of course, the theory
of fields which are not single-valued functions is in no way easier to
formulate
than that for singular fields. It is interesting to note further  that
for boost velocities that are subluminal, one
cannot obtain a globally defined potential $A'^{\mu}$ in any gauge.

We would like to make a few more remarks at this point.
First of all, if we choose another Lorentz frame in which the electric
charge is lightlike while the monopole is moving slowly,
we would get
identical results for the scattering amplitude.  The easiest way to see
this is to use the {\it dual}
formalism wherein one introduces a gauge potential $A_{\mu}^M$
such
that the dual field strength ${\tilde F_{\mu \nu}} \equiv
\partial_{\mu} A_{\nu}^M - \partial_{\nu} A_{\mu}^M $.
If this gauge potential is used to
define
electric and magnetic fields,
then the standard
field
tensor $F^{\mu \nu}$ must satisfy a Bianchi identity of the form
$\partial_{\mu} F^{\mu\nu}=0$
which would then imply that
the gauge potential due to a point charge must have a Dirac
string singularity. Further, the monopole will behave
identically to
the point charge of the usual formalism, so that our method
above is
readily adapted to produce identical consequences. Second, one
can also
treat the scattering of two Dirac monopoles in the same
kinematical limit exactly as in section (\ref{subsection:charge}),
using this dual formalism. This would yield a
result identical to the one for the electric charge case,
with $e$ and $e'$ being replaced by $g$ and $g'$, the monopole
charges. Finally, having dealt with particles carrying either electric
or magnetic charge, it is straightforward to extend our
calculations when one of them is a dyon, that is, it has
both electric and  magnetic charge. The
electromagnetic fields on the
shock front of the
boosted dyon will be the superposition of the fields
produced by a fast charge and a monopole.
Also, depending upon
the nature of the charge on the other particle (electric or
magnetic),
one must employ the usual or the dual
formalism.

With the above observations we are in a position
to address the problem of dyon-dyon scattering in this
formalism. Consider two dyons $\left(e_1,g_1\right)$ and
$\left(e_2,g_2\right)$, where the ordered pair denotes its electric and
magnetic charge contents respectively. Let us assume that the first one
is ultra relativistic. By means of an electromagnetic duality
transformation we can `rotate' a dyon by an angle $\theta$, so that
the new values of electric and magnetic charges become $e'$ and $g'$.
In terms of the old quantities they can be expressed in matrix notation
as
\begin{equation}
\left( \matrix{e' \cr g'} \right) = \left( \matrix{\cos \theta & \sin
\theta \cr -\sin \theta & \cos \theta } \right) \left( \matrix{e_1 \cr
g_1} \right)
\end{equation}
and
\begin{equation}
\left( \matrix{e \cr g} \right) = \left( \matrix{\cos \theta & \sin
\theta \cr -\sin \theta & \cos \theta } \right) \left( \matrix{e_2 \cr
g_2} \right) .
\end{equation}
Now, physical observables do not depend on the parameter $\theta$. We
can make use of this symmetry and choose it to be such that
\begin{equation}
\tan \theta = {g_2 \over e_2}.
\end{equation}
This implies that the first dyon transforms to
\begin{eqnarray}
e'&=&{e_1e_2 + g_1g_2 \over \sqrt {e_2^2 + g_2^2}}  \nonumber  \\
g'&=&{-e_1g_2 + g_1e_2 \over \sqrt {e_2^2 + g_2^2}}
\end{eqnarray}
while for the second dyon
\begin{eqnarray}
e&=&\sqrt {e_2^2 + g_2^2}  \nonumber  \\
g&=&0 .
\end{eqnarray}
This shows that the slow test dyon has been rotated to a pure electric
charge. Then from the results derived previously, the total phase shift
in its wavefunction after being hit by the shock wave of the dyon
$\left( e', g' \right)$ is $ \left[ ee'
\ln {\mu}^2{r_\bot}^2 + 2eg'
\phi \right] . $
Having found this, we can express this in terms of the parameters of the
two dyons we started with. The result is $ \left[ \left( e_1e_2 + g_1g_2
\right) \ln {\mu}^2{r_\bot}^2  - 2\left( e_1g_2 - g_1e_2 \right) \phi
\right] .$
The calculation of the scattering amplitude now
becomes straightforward. It may be noted that the quantities
$ \left( e_1e_2 + g_1g_2
\right)$ and $\left( e_1g_2 - g_1e_2 \right)$ are the only combinations
of the electric charges $e_1,e_2$ and the magnetic charges $g_1,g_2$
that are invariant under duality rotations \cite{sch}. Thus it is
remarkable that the total phase shift and hence the scattering amplitude
depends only on these combinations. Alternatively, we could also have
made the choice  $ \tan \theta = -e_2 / g_2 $, in which case $e$ would
become zero and the second dyon transforms into a monopole. Obviously
these different choices are merely for convenience and the scattering
amplitude does not depend on it. Thus dyon-dyon scattering can always be
reduced to dyon-charge or dyon-monopole scattering. Also note that by a
duality rotation the
usual Dirac quantization condition gets transformed into the generalized
expression
$$ e_1g_2 - e_2g_1 = {n \over 2}   .$$ This implies that the
second term in the phase shift becomes $n \phi$ as in the
charge-monopole scattering case.

Finally, we can ask the question as to what happens if we consider a
massive vector field e.g. that described by the Proca Lagrangian
\begin{equation}
{\cal L}~=~-{1 \over 4}F_{\mu \nu}F^{\mu \nu} + {{\mu}^2 \over 2}
A_{\mu}A^{\mu}
\end{equation}
The solution in the static limit for $A^{\mu}$ in the Lorentz gauge is
given by
\begin{equation}
A^{0}~=~{e'\exp\left(-\mu r\right) \over r}~,~~~~~~~A^{i}~=~0,
\end{equation}
where $e'$ is a point charge at rest. Formally we can apply a Lorentz
boost to this potential and try to take the limit $\beta \rightarrow 1$.
The result is
\begin{equation}
{}^{\beta}A^{\mu}~=~{\eta}^{\mu}{e'\exp \left(-\mu R_{\beta}/{\sqrt {1 -
{\beta}^2}}\right) \over R_{\beta}}
\end{equation}
which vanishes identicaly when we take the limit $\beta \rightarrow 1$.
Thus no shock wave emerges in this case and there are no
$\delta$- function electromagnetic fields on the null plane $x^-=0$. This
observation can also be understood as follows. In the formulation of the
boundary field theory in section (\ref{subsection:scale}) it was shown
that the gauge parameter $\Omega \left({\Omega}^+,{\Omega}^-\right)$ was
the only dynamical degree of freedom in the theory and the corresponding
equations of motion yielded the shock wave picture. On the other hand,
the Lagrangian of the massive vector field does not have the required
gauge invariant structure to admit of such a parameter.
This accounts for the absence of the shock wave.

\section{ELECTROMAGNETIC VERSUS GRAVITATIONAL SCATTERING AT
PLANCKIAN ENERGIES}
\label{section:emgr}

\subsection{Gravity at Planckian energies}

At Planckian c.m. energies the Einstein action also undergoes a
truncation akin to the electromagnetic situation \cite{ver2}. We briefly
sketch how this comes about before summarizing results on the shock wave
geometry and gravitational scattering.  We define the Planck length $l_{pl}$
to be
the inverse of the Planck energy. If $s \gg t$, then the longitudinal
momenta determined by the center of mass energy $\sqrt s$ is much
higher than the typical transverse momenta which
depends on $t$.
Now, if $\sqrt s \approx M_{pl}$, then
correspondingly, the characteristic length scales
associated with the longitudinal direction $l_{\|} \approx l_{pl}$,
while the transverse length scales $l_{\perp} \gg l_{pl}$. We would
also take the coordinates to be dimensionless, in which case
the metric tensor $g_{\mu \nu}$ assumes the dimensions of
{}~$(length)^2$.
With an appropriate coordinate choice, the metric tensor may be cast into
the form
\begin{equation}
g_{\mu \nu}~=~\left(\matrix{g_{\alpha \beta} ~~~0\cr
                       ~~0~~~ h_{ij}}\right)~~.
\end{equation}
Now we make the ansatz that only those
components of the metric become physically relevant which are of
the same order of magnitude as the typical length scales of the
system. In other words, $g_{\alpha \beta} \sim {l_{\|}}^2$ and
$h_{ij} \sim {l_{\perp}}^2$.
If we define two dimensionless metrics $\hat g_{\alpha
\beta}$ and $\hat h_{ij}$ such that
\begin{eqnarray}
g_{\alpha \beta}~=~{l_{\|}}^2~{\hat g_{\alpha \beta}}
\nonumber \\
h_{ij}~=~{l_\perp}^2~{\hat h_{ij}}~~,
\label{eq:gh}
\end{eqnarray}
then it follows that $\hat g$ and $\hat h$ are of the order of
unity.
With these assumptions the usual
Einstein action
\begin{equation}
S_{E} \left[g\right]~=~-{1 \over G} \int d^4x~\sqrt g~R
\end{equation}
splits up into two parts
in the following form,
\begin{equation}
S_{E}\left[g\right]~=~S_{\|}\left[g,h\right]+S_{\perp}\left[h,g\right],
\end{equation}
where
\begin{equation}
S_{\|}\left[g,h \right]~=~-{1 \over G}\int \sqrt g \left(\sqrt h~R_h +
{1 \over4}
\sqrt h h^{ij} \partial_ig_{\alpha \beta}\partial_jg_{\gamma
\delta}{\epsilon}^{\alpha \gamma}{\epsilon}^{\beta \delta}
\right),
\end{equation}
and
\begin{equation}
S_{\perp}\left[h,g\right]~=~-{1 \over G}
\int\sqrt h
\left( \sqrt g~R_g + {1 \over 4}\sqrt g
g^{\alpha \beta} \partial_\alpha h_{ij}\partial_\beta h_{kl}
{\epsilon}^{ik}{\epsilon}^{jl} \right).
\end{equation}
It can be shown that substitution of equation (\ref{eq:gh}) in
the above gives the relation
\begin{eqnarray}
S_{\|}\left[g,h\right]~=~({l_{\|}}/{l_{pl}})^2
{}~S_{\|}\left[\hat g,
\hat h \right], \\
S_{\perp}\left[h,g\right]~=~({l_{\perp}}/{l_{pl}})^2
{}~S_{\perp}\left[\hat h,
\hat g \right]
\end{eqnarray}
where we have used $G \sim {l_{pl}^2}$. From the length estimates
made earlier, we see that the $S_{\|}$ part of the action is
strongly coupled with coupling constant
$g_{\|}={\left(l_{pl}/l_{\|}\right)}^2$
whereas the $S_{\perp}$
part has a weak coupling
$g_{\perp}={\left(l_{pl}/l_{\perp}\right)}^2$.
This shows us that as
far as the transverse directions are concerned (governed by
$g_{\perp}$), the physics is essentially classical, due to the
weak coupling. In fact, the partition function is dominated by
configurations for which $S_{\perp}=0$. It can be shown that here too one
gets a zero curvature constraint,
\begin{equation}
R_{+-}~=~0.
\end{equation}
Once again we are able to justify using a semi-classical method to deal
with such a situation, with the strongly-coupled part of the action
$S_{\pl}$ being treated exactly.

\subsection{Spacetime around a lightlike particle}

The spacetime geometry that emerges for a particle boosted to velocities
close to luminal, is expected to emerge from the coupling of the above
truncated action to a suitably constrained matter energy-momentum tensor.
This has been done in ref. \cite{ver2}. Identical answers can however be
obtained by a process of {\it boosting} the static (Schwarschild) metric
due to
a point particle, adopted in ref. \cite{thf}; we sketch this approach
below. Essentially this boosting means the  mapping of a solution of
Einstein equation with a lightlike particle, namely Minkowski space,
to another Minkowski space but with one of
the null coordinates shifted non-trivially, now without any lightlike
particle present \cite{dray}. It is argued below how this can be
interpreted as a gravitational shock wave.

Once again we choose to carry out
the analysis in a Lorentz
frame in which the velocity of one particle is very much greater than
that of the other. We know that the space time around a point
particle is spherically symmetric and is described by what is
known as the Schwarzchild metric.
If we assume the mass $m$ of the particle to be small, then it is
given in the Minkowski coordinates ($T,x,y,Z$) by,
\begin{equation}
ds^2~=~-\left(1-{2Gm \over R}\right)~dT^2 + \left(1+{2Gm \over R}
\right)
{}~\left(dx^2+dy^2+dZ^2 \right).
\label{eq:sc}
\end{equation}
where $R=\sqrt {x^2+y^2+Z^2}$ and $m \ll R/G$ \cite{dray}.
If the above coordinate
system is moving with a relative velocity $\beta$ with respect to
coordinates $\left(t,x,y,z \right)$ then the two are related by a
Lorentz transformation of the form
\begin{eqnarray}
T~&=&~t\cosh \theta - z\sinh \theta,     \nonumber \\
Z~&=&~-t\sinh \theta+z\cosh\theta,
\label{eq:transf}
\end{eqnarray}
$\theta$ is called the rapidity which is related to the
boost velocity by the relation
\begin{equation}
\tanh \theta~=~\beta.
\end{equation}
Now to take the limit $\beta \rightarrow
1$ or alternatively $\theta \rightarrow \infty$, we also set
\begin{equation}
m~=~2p_{0}~e^{-\theta},
\end{equation}
where the rest energy of the particle is $2p_{0}>0$.
This parametrization is consistent with the fact that
the mass of the particle must exponentially vanish as its velocity
approaches that of light. When we substitute equation ~(\ref{eq:transf})
in equation~({\ref{eq:sc}), we have the metric due to a
particle moving at the speed of light in the $x^{+}$-direction
(i.e along $x^{-}=0$). In terms of the lightcone and the
transverse coordinates this metric becomes
\begin{equation}
ds^2~=~\left(1+{2Gm \over R} \right)\left[-dx^{-}~dx^{+} + dx^2
+dy^2\right] + {4Gm \over R}\left[{p_{0}  \over m}dx{-} +
{m \over {4p_{0}}}dx^{+}
\right]^2,
\end{equation}
with
\begin{equation}
R^2~=~x^2 + y^2 + \left( {p_{0} \over m}x^{-} - {m \over {4p_{0}}}x^{+}
 \right)^2. \label{eq:R}
\end{equation}
Using this and neglecting terms of order $m$ or above, we
get the limiting form of the metric
\begin{equation}
\lim_{m \rightarrow 0}~ds^2= -dx^{-} \left(dx^{+} -
4Gp_{0} {dx^{-} \over
|x^{-}|} \right ) + dx^2 +dy^2 ,
\end{equation}
where the limit is evaluated at $x^{-} \neq 0$ and ($x^{+},x,y$) fixed.
Defining a new set
of coordinates through the relation
\begin{eqnarray}
dx'^{+}~=~dx^{+}-
{4Gp_{0}~dx^{-} \over
|x^{-}|},  \nonumber   \\
dx'^-~=~dx^-  \\
dx'^i~=~dx^i ~, \nonumber
\end{eqnarray}
we observe that the above metric is just a flat Minkowski
metric
\begin{equation}
ds^2 = -dx'^{-}~dx'^{+} + dx'^2 + dy'^2.  \label{eq:met}
\end{equation}
The crucial point to note here is that the metric suffers a
discontinuity at $x^{-}=0$ through the term ${|x^{-}|}^{-1}$.
Now, taking the leading order terms in equation
(\ref{eq:R}), it can be shown that $dx^{-}/x^{-}=dR/R$, which gives
\begin{equation}
dx'^{+}=dx^{+}- \theta (x^{-})~{4Gp_{0}~dR \over R}
\end{equation}
A solution of the above equation near the null plane ( $|x^-|
\rightarrow 0$ ) is,
\begin{equation}
x'^+~=~x^{+} + 2Gp_{0}~ \theta (x^{-})
\ln \left( {\mu}^2 {r_\perp}^2 \right) .
\label{eq:shift}
\end{equation}
Note that the coordinates $x^-$ and $x^i$ remain unchanged.
This step function at the null plane $x^{-}=0$ is the gravitational
equivalent of the electromagnetic shock-wave. There we had a
similar discontinuity in the gauge potential $A^{\mu}$. Here
we have two flat regions of space-time
corresponding to $t<z$ and $t>z$ which are glued together at the
null plane $t=z$ (or~$x^{-}=0$). However there is a shift of
coordinates at this plane given by equation (\ref{eq:shift}). It is as
if a two dimensional flat space-time on the $t-z$ plane is cut
along the line $t=z$ and pasted back again after being shifted along this
line by the amount given above.

Now that we have found the metric around a lightlike particle, in
principle we should be able to predict the behavior of another
(slower) test particle encountering it. Since the
sole effect of the gravitational shock wave is the cutting and pasting of
the Minkowski space along the null direction $x^-=0$ after a shift of the
$x^+$ coordinate, it is easy to see that the test particle wave function
will acquire a phase factor upon passing through this shock front. One
more remark is in order at this point. The
logarithmic singularity in the expression for the shift in the coordinate
$x^+$ in equation (\ref{eq:shift}) causes an infinite time delay of all
interactions via virtual particle exchanges. This shows that it is the
shock wave interactions which dominate over all standard field theoretic
effects like particle creation via brehmstrahlung etc. However, as we
shall show later, the gravitational shock wave may not dominate in all
situations where other interactions mediated also by shock wavefronts exist.

\subsection{Gravitational scattering}

To begin with we will assume the particles to be neutral and as
before, also spinless. We look at the behavior
of the wavefunction of a slow test
particle in the background metric of the lightlike particle
carrying with it a `gravitational' shock wave.
Before the arrival of the shock wave ($x^{-}<0$),
the test particle is in a
flat space time as derived in the last section. Thus, as before,
 its quantum
mechanical wave function is a plane wave of the form
\begin{equation}
\psi_<~(x^{\pm}, {\vec r}_{\perp})~=~e^{ipx}
\end{equation}
with definite momentum $p^{\mu}$.
This can be written in terms of the lightcone and transverse
coordinates as
\begin{equation}
\psi_<(x^{\pm}, r_{\perp})~
=~\exp i \left[ p_{\perp} x_{\perp} - p_{+}x^{-} - p_{-}x^{+}
\right ]
\end{equation}
On encountering the shock wave, it is transported to another
flat space time defined by $x^{-}>0$ which is related to the
previous one by a shift in the $x^{+}$ coordinates.
{}From the explicit expression for this shift in equation
(\ref{eq:shift})
we see that the
wavefunction immediately gets modified into
\begin{equation}
\psi_>(x^{\pm},
{\vec r}_{\perp})~=~\exp i \left[ p_{\perp} x_{\perp} - p_{-} \left( x^{+} +
2G p_0 \ln {r_\perp}^2 \right) \right],
\end{equation}
which is also a plane wave but in the new coordinates.
We have put $\mu = 1$ in equation
(\ref{eq:shift}) and evaluated the above
at $x^{-}=0^{+}$.
Noting that the factor $2G p_{-} p_{0}$
can be written as $Gs$, the phase shift in the final
wave function is
$-Gs \ln r^2_{\perp}.$
But this is just
the electromagnetic phase shift that we got in the last section
in the case of charge-charge scattering with $Gs$ replacing
the earlier coupling $ee'$. This implies that
the scattering amplitude will
also be the same as the previous case with this replacement.
Consequently we have for the gravitational scattering of the two
particles,
\begin{equation}
f(s,t)= {k_+ \over 4\pi k_0}~{\delta(k_+ - p_+)}
{\Gamma
(1-iGs) \over \Gamma (
iGs)}\left (4 \over
-t \right
 )^{1-iGs}~~. \label{eq:scat}
\end{equation}
The corresponding cross section is
\begin{equation}
{d^2\sigma \over d\vec k_{\perp}^2}~ \sim~ {G^2s^2 \over
t^2}
\end{equation}
Despite the striking similarity with electromagnetism, there is an
important difference here. The coupling is now proportional to $s$, the
square of the center of mass energy. The above cross section seems to
increase without limit
with increase of $s$, thus violating unitarity. To understand this, we must
note that at super-Planckian energies one expects gravitational collapse
and inelastic processes to take place. Hence the
above expression fails to be a faithful representation of the
actual scattering and one has to invoke a full
theory of quantum gravity at such extreme energies \cite{thf}.
Similar arguments hold good for all the other cross
sections found in this paper.

Another important point to note is the structure of poles in the
scattering amplitude (\ref{eq:scat}). It seems that there is a `bound
state'  spectrum at
$$ Gs = -iN ~~,~~~~~ N = 1,2,3,\ldots~~~~~~.$$
It has been remarked in \cite{thf2} that the $t$- dependence of the
residues of the poles can be expressed as polynomials in $t$ with degree
$N-1$. Thus, the largest spins of the bound states are $N-1$. This is
similar to the Regge behavior of hadronic resonances, albeit with an
imaginary slope. It remains
to be seen whether these poles are `physical' in the sense they
correspond to resonant states or as argued in \cite{ver2} are just
artifacts of our kinematical approximations.  Nevertheless,
we will show
in the subsequent sections that the introduction of electromagnetism do
have an effect on their location in the complex $s$
plane.\footnote{Unlike in QED, this approximation in gravity cannot be
improved perturbatively because $Gs \sim 1$. For attempts in three and
four dimensions towards improving the `gravity eikonal' see \cite{des},
\cite{amat}, \cite{ven}}.

\subsection{Charge-charge versus gravitational scattering}

After having considered the pure gravitational scattering, we
introduce electromagnetic interactions in the following way.
In addition to their mass, we
now assume the particles to carry electric charges $e$ and $e'$,
$e$ being the charge of the slow test particle.
Then the  charge $e'$ will also have an electromagnetic shock wave
associated with it. The electric and magnetic fields on the shock
front are those found in the previous section, given in equation
(\ref{eq:efield}).
We assume that the resultant effect of the combined shock wave
(gravitational and electromagnetic) on the test particle is to
produce a phase shift in its wave function which is the sum of
the individual phase shifts. This tacitly presumes the independence of
the gravitational and
electromagnetic shock waves. We shall not attempt to prove this
supposition at this point except to note that this assumption has also
been made without explicit mention in previous works
\cite{jac,thf,das}.
However, it can be justified rigorously for a variety of
situations \cite{dm}.
Both the phase shifts being proportional to $\ln {\mu}^2r_\perp^2$, the
net effect is succintly captured by the shift $Gs \rightarrow Gs+ee'$, with
the final form of the wave function
after is crosses the null plane $x^{-}=0$ being
\begin{equation}
\psi _>(x^{\pm},x_{\perp})
{}~ =~\exp~\left[-i\left(ee^{'}+Gs\right)\ln\mu^{2}r^{2}_{\perp} +
ipx\right]~.
\end{equation}
Consequently, the scattering amplitude becomes
\begin{equation}
f(s,t)= {k_+ \over 4\pi k_0}~{\delta(k_+ - p_+)}
{\Gamma (1-iee'-iGs) \over \Gamma (iee'+iGs)}\left (4 \over-t \right
)^{1-iee'-iGs}~~. \label{eq:eegr} \end{equation}

\noindent
This gives the cross section,
\begin{equation}
{d^2\sigma \over d\vec k_{\perp}^2}~ \sim~ {1
\over t^2} \left ( ee'
+ Gs \right )^2~~. \label{eq:emgrcsec}
\end{equation}
To compare the relative magnitudes of the two terms, we recall
that the electromagnetic coupling constant $ee'$ evolves only
with $t$ through radiative corrections and not with $s$. Thus in
the kinematical regime that we are considering, it remains fixed
at its low energy value. For example, if the particles
carry one electronic charge each, then $ee' \sim  1/137$.
On the other hand, at Planck scales, the second term in the cross
section is of order unity. This shows that gravity is the
principal contributor in the scattering process and
electromagnetic effects can be treated as small perturbations. Likewise,
the poles of the scattering amplitude (\ref{eq:eegr}) are shifted by ${\cal
O}(\alpha)$ corrections to the pure gravity poles. Observe that these
poles appear only when gravitational interactions are taken into account,
because it is only in this case that the interaction is a (monotonically
increasing) function of energy.

\subsection{Charge-monopole versus gravitational scattering}

Motivated by the conclusions of the last section, we now
proceed to investigate whether they undergo any modifications
when we assume one of the particles to carry a magnetic
charge. In other words, will gravity still dominate
over electromagnetic interactions at Planckian energies? With the
replacement of the electric charge $e'$ of the fast moving particle by a
magnetic charge $g$, the fields on the electromagnetic
shock front are given by equation (\ref{eq:fld}).
As before, when it crosses the charge $e$, we add the
gravitational and electromagnetic phase shifts in its
wavefunction. While the former is still $-Gs\ln {r_{\perp}}^2$, the latter,
as seen from equation (2.39), is now $in\phi$. Thus, charge-monopole
electromagnetic effects cannot be incorporated by a shift of $Gs$, in
contrast to the charge-charge case. Thus the wavefunction assumes the form
\begin{equation} \psi _>(x^{\pm}, x_{\perp})~ =~
\exp~\left[i\left(n\phi - Gs\ln\mu^{2}r^{2}_{\perp} +
ipx \right) \right]
\end{equation}
Due to the azimuthal dependence, the calculation of the
overlap with momentum eigenstates has to be done ab initio.
Clearly, the relevant integral for the evaluation of
$f(s,t)$ is
$$\int d^2 r_{\bot}~\exp[i(n\phi - Gs~
\ln{ \mu ^2 r_{\bot}^2}  +\vec q . \vec r_{\bot})].$$
Once again, the integral over $\phi$ is readily done, and
the above reduces to
\begin{equation}
{1 \over q^2} \int_0^{\infty} d \rho ~\rho^{1-2iGs} J_n(\rho)~.
\end{equation}
Here $J_n(\rho)$ is the Bessel function of order $n$. The
above integral is again a standard one \cite{luke} and
finally we get the amplitude
\begin {equation}
f(s,t)= {k_+ \over 4\pi k_0}~{\delta(k_+ - p_+)}~\left({n\over
2}-iGs\right){\Gamma
({n \over 2}-iGs) \over \Gamma ({n \over 2} + iGs)}\left (4 \over
-t \right
)^{1-iGs}~~ \label{eq:ram}
\end{equation}
and hence the cross section
\begin{equation}
{d^2\sigma \over d\vec k_{\perp}^2}~ \sim~ {1 \over t^2} \left (
{n^2 \over4}
+ G^2s^2 \right )~~. \label{csec}
\end{equation}
Since $n$ is at least of order unity,
it is clear from the above expression, that for $\sqrt s
\approx M_{pl}$, both the terms are of the same order of
magnitude. This means that unlike charge-charge scattering,
even at Planck scale gravity is no longer the dominant shock wave
interaction. Electromagnetism with monopoles becomes
equally important. This dramatic difference from the charge-charge case
is a consequence of the Dirac
quantization condition, which restricts the values of $e$
and $g$ from being arbitrarily small. In fact, the above may be
considered to be a
rephrasal of the strong coupling aspects of the monopole sector in
electromagnetism and of the gravitational interactions at Planck scale.
As already mentioned earlier, gravitational effects would indeed tend to
dominate for $Gs >>1$ if the Dirac quantum number $n$ is held fixed. But
it is far from clear if, in this circumstance, the simple-minded
semiclassical analysis performed above will go through without
modification. Indeed, as explained in ref. \cite{thf}, super-Planckian
energies will most probably entail real black hole collisions with the
ensuing technical complications.

Returning once more to the analytic structure of $f\left(s,t\right)$, we
see that now they occur at
$$Gs = -i\left( N + {n \over 2 } \right)~~, $$
that is a shift in $s$ by half-odd integral values. Once again, the
spectrum of these `bound states' is no longer a perturbation on the
spectrum in the pure gravity situation. More interestingly,
notwithstanding claims in the literature (cf. \cite{ver2})  that the 't
Hooft poles are
artifacts of the large impact parameter approximation, the shift observed
above due primarily to the monopoles strongly suggest another
possibility: the {\it Saha phenomenon} \cite{saha}. Recall that, this
implies that any charge-monopole pair composed of spinless particles
will, as a consequence of Dirac quantization, possess half-odd integral
quantized (field) angular momentum. If we blithely regard the integer $N$,
which also occurs in the spectrum of bound states in pure gravitational
scattering, as the {\it spin} of the states, then it is enticing to
consider the shift by one-half
the Dirac quantum number $n$ in the charge-monopole case to be the extra
spin that the system would pick up in accord with Saha's predictions.
Further, if one speculatively associates the Regge-like behavior observed
in purely gravitational scattering with the spectrum of some string
theory (albeit with imaginary slope parameter), then the spectrum with
charge-monopole electromagnetic scattering
can as well be speculated to correspond to some {\it supersymmetric}
string theory. In any event the role of electric-magnetic duality, were
we to actually discern any such string structures, can hardly be
over-emphasized.

\section{CONCLUSION}

While reinforcing the general result that at c.m. energies of the order
of the Planck scale and low momentum transfer, two particle
scattering is primarily a shock wave phenomenon with standard exchange
processes relegated to relative unimportance, our work emphasizes the
role of electromagnetic shock waves associated with the magnetic
monopole sector. Since this sector is generically a strong coupling one
akin to gravity at Planckian energies, it is not surprising that the
contributions of the two interactions to the cross section are
comparable. While similar cross sections have been computed for gravity
within string theories \cite{amat} which are ostensibly correct theories
of quantum gravity with tractable ultra-violet behavior, it will be
interesting to see if the recently proposed `dual' strings \cite{sen}
(or some modification thereof) exhibit the behaviour observed above. The
major advantage of the shock wave picture is its universality in dealing
with gauge particle exchanges within  this, albeit somewhat restricted
kinematical region. Even when a well-defined local field theory is not
available, non-trivial physical information can indeed be obtained within
this picture. The task that remains then is to formulate the theory in
such a way that a systematic procedure is available to compute
corrections to the predictions given by this picture \cite{des},
\cite{amat}, \cite{ven}.

The assumption of decoupling of electromagnetic and gravitational shock
waves that we have made above, of course warrants justification, even
though similar assumptions have been tacitly made in earlier work. This
decoupling will be crucial if one wishes to apply the shock wave picture
to analyze gravitational collapse and Hawking radiation from black holes
\cite{thfbh}, where the relevant particles carry electric and magnetic
charge, or charged particles scatter off charged black holes.
It appears that such decoupling happens quite naturally for scattering of a
class of particles whose electromagnetic and gravitational fields are
derived from those of Reissner-Nordstrom black holes. However, for
dilatonic charged black holes, the corresponding shock waves exhibit some
mixing which may have novel physical consequences. We hope to report on this
in a future publication \cite{dm}.

\vspace{.7cm} \begin{center} {\bf \large {Acknowledgements}} \end{center}

We thank R. Anishetty, G. Date, T. Jayaraman, G. Rajasekaran, P. Ramadevi,
T. Sarkar, H. Sharatchandra and C. Srinath for useful discussions. We
also thank R. Basu for a careful reading of the manuscript.

\end{document}